\documentclass{aastex}          
\usepackage{spr-astr-addons}   

\begin{document}
%
\title{Brane viscous cosmology in the plasma era}

\shorttitle{Brane viscous cosmology }
\shortauthors{Brevik }

\author{Iver Brevik\altaffilmark{1}}
\affil{Department of Energy and Process Engineering, Norwegian University of Science and Technology, Trondheim, Norway}
\email{iver.h.brevik@ntnu.no} 


\begin{abstract}

We consider how the five-dimensional Randall-Sundrum (one-brane) theory becomes modified when account is taken of the bulk viscosity of the cosmic fluid on the brane. We focus on the plasma era between $10^{12}~$K (muon pair annihilation) to about $5\times 10^9~$K (electron-positron annihilation), which includes the first order quark-hadron transition beginning at an energy density of about $5\times 10^9~\rm MeV^4$.  Various possibilities are examined for modeling the bulk viscosity, preference being at the end given to the results calculated from relativistic kinetic theory. According to this, the viscosity is negligible at the highest temperatures, but may amount to a few per cent corrections in the later stages of the plasma era. We also briefly consider   anisotropic universes where the shear viscosity comes into play, and show that in the case of the Kasner model the influences from  bulk viscosity and  shear viscosity become comparable when the anisotropy parameter of the universe is of order $A \sim 10^{-11} $ in the beginning of the plasma era, and $A \sim 10^{-2}$ in its later region.

\end{abstract}


\section{Introduction}

Up to now, various holographic approaches to the ${\cal N}=4$ supersymmetric
Yang-Mills (SYM) theory with strong coupling
have been performed from the dual supergravity viewpoint; cf.
 \cite{MGW,KK,KMMW,KMMW2,Bab,ES,SS,NPR,GY,CNP}.
In these approaches, the research has been extended
to the SYM theory
in the background of dS$_4$ (AdS$_4)$ by introducing the 4D cosmological constant
($\Lambda_4$); cf. \cite{Hawking,Alishahiha1,Alishahiha2,H,GIN1,GIN2,EGR,EGR2}.
In this way it has been found that
the dynamics of the SYM theory is controlled by
the 4D geometry, especially for the dS$_4$ and AdS$_4$ cases; cf. \cite{GIN1,GIN2}.
In this kind of theory, the 4D geometries  in cosmology are used.

\vspace{.5cm}
\noindent  Now, one might  use the results of this holographic SYM to solve the field equations in
 4D cosmology. A characteristic properties of such a method is that it is
 equivalent to investigating the back reaction from SYM theory on 4D cosmology. We intend to analyze this topic in a later work. In the present paper we will however focus attention on another, though related aspect of the problem, namely brane world cosmology in the early universe. Specifically, we will consider the  Randall-Sundrum theory for a single brane situated in AdS$_5$ bulk space, and concentrate on the lepton era for which the temperature drops from about $10^{12}~$K (muon pair annihilation) to about $5\times 10^9~$K (electron-positron pair annihilation). The universe is then filled with electrons, three types of neutrinos, the corresponding antiparticles, and photons; cf.  \cite{hogeveen86}.

 A key reference for us will be the work of  \cite{derisi08}, dealing with the first order quark-gluon phase transition in  brane-world cosmology. We will generalize their analysis by taking into account the {\it bulk viscosity} in the cosmic fluid. That such an investigation is natural to undertake, follows from the fact that the viscosity coefficients ($\eta$ for the shear viscosity, $\zeta$ for the bulk viscosity) in the plasma era are very large. At least nominally, this is so. In particular, it is evident for the shear viscosity: under ordinary laboratory circumstances $\eta$ is of order $0.01~\rm g~cm^{-1}s^{-1}$ (water), but in the beginning of the lepton era it is a factor of about $10^{26}$ larger; cf. \cite{hogeveen86}. Also, for the bulk viscosity the difference is very high (a factor of about $10^{15}$ larger). Of course, a main reason for these big numbers is that the energy densities are so high in the early universe. To see the actual influence from viscosity, one has to carry out the calculations explicitly.

 In the next section we first present the general formalism for 5D viscous brane theory, discuss the temperature and the duration of the first order transition, and consider thereafter three different options for how to model the bulk viscosity. Our preferred choice is the expression for $\zeta(T)$ calculated from kinetic theory. Using it, we find that at the highest temperatures in the plasma era the influence from friction is completely negligible. At the lowest temperatures, the influence from $\zeta(T)$ may amount to a few per cent. The general relationship between temperature $T$ and time $t$ is given by Eqs.~(\ref{33}) and (\ref{34}) below, for the post-transitional period where the fluid has entered the hadron phase.

 As usual in cosmology, we will for the most part ignore the shear viscosity; this being   in accordance with the common assumption about spatial isotropy in the cosmic fluid. In recent times there has been an increased interest in anisotropic cosmology, however, for a large  part due to observations. An anisotropic universe immediately brings the shear viscosity into play. In the final section we consider, as a typical example, the anisotropic Kasner universe, calculating the production of specific entropy. It turns out that
  on {\it both} viscosity coefficients, $\eta$ and $\zeta$, contribute, and it is possible to relate the rate of entropy to the anisotropy parameter, usually called $A$, in the universe. We find that at the highest temperatures the value of $A$ necessary to make $\zeta$ and $\eta$ of the same degree of importance, is immensely small. At the lowest temperatures, however, a value of $A \sim 10^{-2}$ will cause the two viscosity coefficients to be about equally important for the entropy production.

It should be mentioned that we neglect the viscosity of {\it dark radiation.} Very little seems to be known about such a viscosity, if it exists at all.

\section{One-brane viscous Randall-Sundrum theory}

\subsection{General formalism}

As mentioned above, we shall consider one single brane situated in five-dimensional AsS$_5$ bulk space. This is the Randall-Sundrum II model; cf. \cite{randall99}, when the fluid on the brane has a bulk viscosity $\zeta$. We shall omit the dS$_5$ case. We begin by recapitulating some essentials of this theory, as were spelled out in \cite{brevik02,brevik04} (cf. also the related \cite{brevik06,brevik08}). Let the brane be located at $y=0$, the surrounding bulk being empty except from the 5D cosmological constant $\Lambda_5$. The metric is taken to have the form
\begin{equation}
ds^2=-n^2(t,y)dt^2+a^2(t,y)\gamma_{ij}dx^idx^j +dy^2, \label{1}
\end{equation}
where  $\gamma_{ij}=(1+k\delta_{mn}x^mx^n/4)^{-2}\delta_{ij}$, and where
$n(t,y)$ and $a(t,y)$ are determined by Einstein's equations
\begin{equation}
R_{AB}-\frac{1}{2}g_{AB}R+g_{AB}\Lambda_5=\kappa_5^2T_{AB}. \label{2}
\end{equation}
The coordinates are $x^A=(t,x^1,x^2, x^3, y)$, with $\kappa_5^2=8\pi G_5$ the five-dimensional gravitational coupling. Einstein's equations are given in the references mentioned above, and will not be reproduced here. The form of the energy-momentum tensor $T_{AB}$ is not so far specified, but as the 5D space outside the brane is assumed empty the components of $T_{AB}$ are different from zero only on the brane.

 On the brane (subscript zero) the tension  $\sigma$ is assumed to be constant. The total energy-momentum tensor can be written as
\begin{equation}
T_{AB}=\delta(y)(-\sigma g_{\mu\nu}+T_{\mu\nu}^{\rm fluid})\delta_A^\mu\delta_B^\nu, \label{3}
\end{equation}
where $T_{\mu\nu}^{\rm fluid}$ is the fluid part
\begin{equation}
T_{\mu\nu}^{\rm fluid} =\rho U_\mu U_\nu+(p-\zeta \theta)h_{\mu\nu}. \label{4}
\end{equation}
Here $U^\mu$ is the fluid's four-velocity ($U^\mu U_\mu =-1$), $h_{\mu\nu}=g_{\mu\nu}+U_\mu U_\nu$ is the projection tensor, and  $\theta=3\dot{a}_0/a_0+\dot{n}_0/n_0$ is the scalar expansion. For simplicity we use the notation $a_0(t)=a(t,y=0)$, $n_0(t)=n(t,y=0)$.

Imposing the gauge condition $n_0(t)=1$ on the boundary we get $\theta=3\dot{a}_0/a_0=3H_0$. The shear viscosity, as mentioned,  can be omitted when we assume spatial isotropy of the fluid. We shall work in the orthonormal frame where $U^\mu=(1,0,0,0)$. The metric on the brane becomes thus
\begin{equation}
ds^2=-n_0^2(t)dt^2+a_0^2(t)\gamma_{ij}dx^idx^j. \label{5}
\end{equation}

 We have to consider the junction conditions across the brane (overdots and primes meaning derivatives with respect to $t$ and $y$ respectively).  The distributional parts of $a''$ and $n''$ must match  the distributional parts of the energy-momentum tensor. We write $a''=\hat{a}''+[a']\delta(y)$, where $[a']=a'(y=0^+)-a'(y=0^-)$ is the jump across $y=0$ and $\hat{a}''$ is the nondistributional part.  Similar expressions hold for $n$. Assuming that there is no flux of energy in the $y$ direction ($T_{ty}=0$), one derives as a consequence that
\begin{equation}
n(t,y)=\frac{\dot{a}(t,y)}{\dot{a}_0(t)}. \label{6}
\end{equation}
From the field equations we find, by integration over $y$,
\begin{equation}
\left( \frac{\dot{a}}{na}\right)^2+\frac{k}{a^2}=\frac{\Lambda_5}{6}+\left( \frac{a'}{a}\right)^2+\frac{C}{a^4}, \label{7}
\end{equation}
where $C$ is an integration constant. On the brane $y=0$ this implies
\begin{equation}
\left(\frac{\dot{a}_0}{a_0}\right)^2+\frac{k}{a_0^2}=\lambda_4+\frac{\kappa_5^4\sigma \rho}{18}+\frac{\kappa_5^4\,\rho^2}{36}+\frac{C}{a_0^4}, \label{8}
\end{equation}
where $\lambda_4$  is the effective 4D cosmological constant,
\begin{equation}
\lambda_4=\frac{\Lambda_5}{6}+\frac{\kappa_5^4\,\sigma^2}{36} \label{9}
\end{equation}
(note that as $\sigma$ is  constant, so is $\lambda_4$). It is to be observed that $\lambda_4$ contains the 5D cosmological constant $\Lambda_5$ as well as the  tension $\sigma$ on the brane. Note also
 that Eq.~(\ref{8}) does not contain the viscosity explicitly.

We ought here to point out that the brane equation (\ref{8}) is formally similar to the corresponding equation  in  holographic cosmology;  the subscript zero in that case to be associated with the AdS boundary instead of with the brane position $y=0$. As already mentioned, we intend to return to this correspondence in a later work.

\bigskip

\noindent Let us  turn to the energy conservation equation in the presence of bulk viscosity,
\begin{equation}
\dot{\rho}+3(\rho +p)H_0-9\zeta H_0^2=0. \label{10}
\end{equation}
 Here the junction condition for $n$ is taken into account; cf.  \cite{brevik04}. This equation is the same as obtained in ordinary 4D viscous cosmology; this may appear  somewhat surprising as there is no obvious reason why it should be so. We see that the viscosity now appears explicitly in Eq.~(\ref{10}).

 Note that Eq.~(\ref{10}) can be understood as the usual continuity equation,
 \begin{equation}
 \dot{\rho}+3(\rho+\tilde{p})H_0=0, \label{10A}
 \end{equation}
 (for  $\tilde{p}=p-\zeta \theta)$, which can be read off from the definition of $T_{\mu\nu}$ given above.

A note on dimensions: It is convenient to let the dimensions be carried  by the coordinates, so that $a_0$ and $n$ become nondimensional.  Thus, in geometric units $[a_0]=[n]=1, [k]=[\lambda_4]=  [C]=    {\rm cm}^{-2}, [\kappa_4]^2={\rm cm}^2, [\kappa_5]^2={\rm cm}^3, [\sigma]=[\rho]={\rm cm}^{-4}, [\zeta]={\rm cm}^{-3}.$ Since it is customary to use natural energy units in this field, one may note the useful conversion formulas
\begin{equation}
 1 {\rm MeV}=5.068\times 10^{10}\, {\rm cm}^{-1}=1.520\times 10^{21}\,\rm {s}^{-1}, \label{x}
 \end{equation}
which imply that $ 1 \rm MeV^4= 2.085\times 10^{26}\, erg\, cm^{-3}. $

\subsection{ Influence from viscosity in the plasma era}

We first make the following  observation: Eq.~(\ref{8}) corresponds formally, term by term, to Eq.~(6) in Ref.~\cite{derisi08}, that paper being concerned with the  quark-hadron phase transition in the lepton era. The last term in their Eq.~(6)  (called $C/a_0^4$ here), was interpreted as the projection of the 5D Weyl tensor onto the brane. (To see the correspondence one needs to note that the relationship between the 4D and 5D gravitational couplings is $\kappa_4^2=\kappa_5^4\sigma /6.$).

We will now investigate how the relationship between cosmic time $t$ and temperature $T$ becomes influenced by the bulk viscosity, both before the phase transition, and after it. It is helpful first to give some data of the transition; cf. \cite{derisi08}:
  It starts at a critical temperature of $T_c=125~$MeV $(1.45\times 10^{12}$ K) when the universe is a few tenths of $\mu$s old. The energy density is then $\rho_Q \approx 5\times 10^9~\rm MeV^4$ ($10^{36} \rm \,erg~cm^{-3}).$ Before this instant, the cosmic fluid is a quark-gluon plasma. During the transition $T_c $ is constant, as is also the critical pressure $p_c=4.6\times 10^8~\rm MeV^4$. At the end of the transition $\rho(t)$ is decreased to $\rho_H \approx 1.38\times 10^9\, \rm MeV^4$ $(2.87\times 10^{35}\, \rm erg~cm^{-3})$. The universe enters then  the hadron phase.
We can calculate the duration $\Delta t=t_h-t_c$ of the transition by making use of the formalism of Ref.~\cite{derisi08}. A simple calculation gives $\Delta t \approx 120~\mu$s, when the brane tension $\sigma$ is taken to have the typical value
\begin{equation}
\sigma=5\times 10^{10} \, \rm MeV^4. \label{13}
\end{equation}
This value of $\Delta t$ is relatively large. The physical reason why the new phase does not turn up immediately is that, as a characteristic property of a first order transition, some supercooling is needed to overcome the energy expense in forming the new phase. (Note: the viscosity is not included in this calculation of $\Delta t$.)

Let us next consider the equation of state for the fluid. In the initial quark phase we write it  in the form
\begin{equation}
\rho_q=3a_qT^4+B, \quad p_q=a_qT^4-B, \label{14}
\end{equation}
where; cf.  \cite{derisi08}, $a_q=(\pi^2/90)g_q,$ with $g_q=16+(21/2)N_F+14.25=51.25$ and $N_F=2.$
 The $B$ is the bag constant, whose numerical value is usually given by  $B^{1/4} \approx  200$ MeV. There exist also two temperature corrections to Eq.~(\ref{14}), one term proportional to $T^2$ and the other proportional to, but they are expected to be small and will be neglected here for simplicity.

We now need to model the bulk viscosity $\zeta$ in Eq.~(\ref{10}) (cf., for instance, \cite{gron90} and \cite{brevik13}).  This can be done in various ways. To begin with, we might follow the ansatz of Murphy in his classic  paper on viscous cosmology (\cite{murphy73}), and assume the "kinematic viscosity" $\zeta/\rho$ to be a constant. Near the Big Bang where $\rho$ is large the same should accordingly hold for $\zeta$; thereafter $\zeta$ is predicted to decrease with time. This is however the opposite of the behavior one calculates from relativistic kinetic theory. So Murphy's ansatz seems to overpredict the influence from viscosity in this era of the universe's history.

As a second option, we might assume the kinematic viscosity to be inversely proportional to   $H_0$, $\zeta/\rho \propto 1/H_0$.
This ansatz was actually  made use of in \cite{brevik04}. An advantage of it was that the mathematical handling of the hydrodynamic formalism was found to be  quite simple. As in the present case the bag constant  is not expected to have any influence on the viscosity, we can
 omit $B$ in the expression (\ref{14}) for $\rho_q$  and substitute only the thermal part $3a_qT^4$  to get for the quark fluid
\begin{equation}
\zeta=\frac{4\beta}{9H_0}(3a_qT^4) \label{15}
\end{equation}
with $\beta$ a nondimensional constant (the factor 4/9 is introduced for convenience).

Since $\rho_q+ p_q=4a_qT^4$ from Eq.~(\ref{14}) we  obtain from Eq.~(\ref{10}), inserting Eq.~(\ref{15}) for $\zeta$, an equation for $\dot{\rho}_q$ that can be  manipulated to give the temperature dependence of the scale factor in a simple form,
\begin{equation}
T(a_0)=\frac{T_*}{a_0^{1-\beta}}, \label{16}
\end{equation}
the asterisk meaning a reference value, corresponding formally to the temperature when $a_0=1$.   If the fluid is nonviscous, $\beta=0$, one gets $T(a_0)=T_*/a_0$, in accordance with \cite{derisi08}, Eq.~(20). For a given value of the scale factor $a_0$ on the boundary $y=0$, the temperature in the viscous case becomes thus modified by a factor $a_0^\beta$.

Let us explore the consequences of the ansatz (\ref{15}) one step further, by evaluating what correspondence between temperature and time it actually predicts.  For definiteness we chose an initial time $t$ prior to the transition temperature $t_c$ (quark-gluon era). We then have to go back to Eq.~(\ref{8}), in which we insert $H_0=\dot{a}_0/a_0=-[1/(1-\beta)]\dot{T}/T$. Integrating from $T(t)$ up to $T_c$ we then get
\begin{equation}
t_c-t=\frac{1}{1-\beta}\int_T^{T_c }\frac{dT}{TD(T)}, \label{17}
\end{equation}
where
\[
D(T)= \Big[ \lambda_4+\frac{\kappa_5^4\sigma}{18}B-k\left(\frac{T}{T_*}\right)^{\frac{2}{1-\beta}} \]
\begin{equation}
+\frac{\kappa_5^4\sigma}{6}a_q T^4
+\frac{\kappa_5^4}{36}(3a_qT^4+B)^2   +  C\left(\frac{T}{T_*}\right)^{\frac{4}{1-\beta}}  \Big]^{1/2}. \label{18}
\end{equation}
 The viscosity thus occurs in the prefactor $1/(1-\beta)$ in the integral (\ref{17}), and in   the spatial curvature term ($k$) as well as in the dark radiation term ($C$) in Eq.~(\ref{18}).

  On physical grounds it is however of interest   to ask:  can we say something  about  the  connection between the ansatz (\ref{15}) and general  expressions for bulk viscosity in hydrodynamics?  Let us recall  in this context  the following formula, derived for photons,
  \begin{equation}
        \zeta = 4a_{\rm rad}T^4 \tau \left[ \frac{1}{3}-\left( \frac{\partial p}{\partial \rho}\right)_n  \right]^2, \label{19}
        \end{equation}
         in which $\tau$ denotes the mean free time. This formula is attributed to  \cite{thomas30}, and is given also in the review article by \cite{weinberg71}. The formula gives us the following physical insight: its first (leading) term  corresponds to the ansatz (\ref{15}) {\it only if $\tau \propto 1/H_0$. }

          In view of this restriction it is perhaps not so unexpected after  all that  the ansatz  (\ref{15}) does not appear to be possessed by the cosmic fluid in the very early universe. Kinetic theory calculations - which we shall in the following consider as the most reliable ones - indicate that at these high temperatures the value of $\beta$ is not a constant but is instead a function that varies strongly with  temperature. It might well right that the ansatz (\ref{15}) is a good one at later stages of the universe's history, but we will further abandon this form  in this paper. So, as our third and final option, we will henceforth simply adopt the results from kinetic theory to the first order in inverse temperature as given by \cite{hogeveen86}:
 \begin{equation}
 \zeta(T)=ET^{-5}, \quad \rm with \quad E=2.8\times 10^{73}\, g\,cm^{-1}s^{-1}K^5,
  \label{20}
 \end{equation}
 (cgs units). Also, we will henceforth focus on late part of the plasma era, which is the only part of it where viscosity seems to be of physical importance at all.

 Assume now temperatures $T \leq 10^{10}\,$K, when the fluid is in the hadronic region. The energy density is then $\rho_h=3p_h$. We write $\rho_h$ in the form
 \begin{equation}
 \rho_h=g_ha_{\rm rad} T^4, \label{21}
 \end{equation}
 where
 \begin{equation}
 a_{\rm rad}=\frac{\pi^2k_B^4}{15\hbar^3c^3}=7.56\times 10^{-15}\, \rm erg\, cm^{-3}K^{-4} \label{22}
 \end{equation}
 is the radiation energy constant, and $g_h=17.25$. Equation (\ref{10}) now reads
 \begin{equation}
 \dot{\rho}_h+4\rho_hH_0-9ET^{-5}H_0^2=0, \label{23}
 \end{equation}
 which can be rewritten as ($H_0=\dot{a}_0/a_0$)
 \begin{equation}
 \frac{\dot{T}}{T}+\frac{\dot{a}_0}{a_0}-\frac{9E}{4g_ha_{\rm rad}T^9}\frac{\dot{a_0}^2}{a_0^2}=0. \label{24}
 \end{equation}
It is of interest to evaluate the ratio between the last two terms in this equation. Calling the ratio $X$, we get
\begin{equation}
X=\frac{9E}{4g_ha_{\rm rad}}\frac{1}{T^9}\frac{\dot{a}_0}{a_0}=4.83\times 10^{86} \frac{1}{T^9}\,\frac{\dot{a}_0}{a_0}. \label{25}
\end{equation}
This expression is useful, since it shows the very high sensitivity of the viscosity with respect to temperature, and also gives a realistic estimate of the magnitude of the viscosity effect. We do not need to solve the full problem in order to estimate the magnitude of this; we may only use the known expressions for the  radiation dominated universe:
\begin{equation}
a(t)=2.2\times 10^{-10} t^{1/2}, \quad T(t)=10^{10}t^{-1/2}\, \rm K, \label{26}
\end{equation}
 according to which  $t=1~$s when $T=10^{10}~$K. Then $H_0=1/(2t)=1/2\, \rm s^{-1}$, and so $X \approx 2\times 10^{-4}$ at this temperature. This the order of magnitude of viscosity correction that we may expect. If we stretch the expansion (\ref{20}) to the lower limit of the plasma era, $T=5\times 10^9~$K $(H_0=1/8~\rm s^{-1})$, we will get $X \approx 0.03$. These numbers are not quite negligible, and indicate that we ought to  treat Eq.~(\ref{24}) a bit further.

 We follow this up by first writing Eq.~(\ref{24}) as
 \begin{equation}
 \frac{d}{dt}\ln (Ta_0)=Kt^{5/2},   \label{27}
 \end{equation}
 where we have made use of Eqs.~(\ref{26}) to express the viscous term as a function of $t$. The constant $K$ is
 \begin{equation}
 K= \frac{9E}{16g_ha_{\rm rad}}\frac{1}{10^{90}}= 1.21\times 10^{-4}, \label{28}
 \end{equation}
 in cgs units. This equation can be integrated with respect to $t$, from the chosen initial instant $t=1~$s to an arbitrary higher value of $t$ at least roughly compatible with the approximation (\ref{20}). At the initial point we moreover assume the validity of Eqs.~(\ref{26}), and we define   for simplicity
 \begin{equation}
 T_* =2.2 \, \rm K, \label{29}
 \end{equation}
 in order to comply with the notation used earlier. We then get from Eq.~(\ref{27})
 \begin{equation}
 a_0(T)=\frac{T_*}{T}\,e^{F}, \label{30}
 \end{equation}
 where
 \begin{equation}
 F=3.45\times 10^{-5}(T_{10}^{-7}-1), \quad    T_{10}=T/10^{10},            \label{31}
 \end{equation}
 (it is helpful to note that $H_0=\frac{1}{2}T_{10}^2).$  As $T_{10} <1$ in the integration domain, we thus see that $F<0$ and $a_0(T)$ becomes slightly smaller than in the nonviscous case.

  With  Eq.~(\ref{30}), implying
  \begin{equation}
  \frac{\dot{a}_0}{a_0} =-\frac{\dot{T}}{T}+\dot{F},  \quad  \dot{F}=KT_{10}^{-5},   \label{32}
  \end{equation}
  we obtain from Eq.~(\ref{8})
  \begin{equation}
  t-1=\int_T^{T_{10}=1}\frac{dT}{T[R(T)-\dot{F}]}, \label{33}
  \end{equation}
  with
 \[  R(T)=\]
 \begin{equation}
  \left[ \lambda_4-k\left(\frac{T}{T_*}\right)^2 e^{-2F}+\frac{\kappa_5^4\sigma}{18}g_ha_{\rm rad}T^4+\frac{\kappa_5^4}{36}(g_ha_{\rm rad})^2T^8     +  C\left(\frac{T}{T_*}\right)^4 e^{-4F}               \right]^{1/2}. \label{34}
  \end{equation}
The integral (\ref{33}) requires numerical calculation. We will abstain from this, in view of the smallness of the effect. Our main purpose has been to give the general method for how to calculate the influence from viscosity; the method of course works also when the viscosity is greater than in the present case. One has only to adjust input formula (\ref{20}) for the viscosity appropriately.

Of obvious physical interest is however to consider the magnitude of the nonlinear $(\rho^2$) term as compared with the linear term in Eq.~(\ref{8}), under the present circumstances.  Calling the ratio $Y$, we see that $Y=\rho/(2\sigma)$. Thus with the same value for the brane tension as used above, $\sigma=5\times 10^9\, \rm MeV^4$, we see that in the quark era
\begin{equation}
Y_Q= 0.05, \label{35}
\end{equation}
whereas in the hadron era
\begin{equation}
Y_h=1.38\times 10^{-2}. \label{36}
\end{equation}
Thus, the influence from  nonlinearity decreases from  5\% to about 1\% during the plasma era.

\section{Remarks on the shear viscosity}

We have so far considered the bulk viscosity $\zeta$ only.  As is known, the shear viscosity $\eta$ is usually omitted in cosmology because of the assumed spatial isotropy of the cosmic fluid. In later years   the  shear viscosity concept has however attracted increased attention, because of its  importance   in different areas of physics. Thus, one may notice that in the modeling of elementary particles this concept plays an essential role in connection with the suggestion   about the existence of   a universal lower bound on the ratio $\eta/s$,  $s$ being the entropy content per unit volume (cf., for instance, the classic  paper by \cite{kovtun03}).  In geometric units the proposed inequality takes the simple form
\begin{equation}
\frac{\eta}{s}>\frac{1}{4\pi}. \label{38}
\end{equation}
There have later appeared several other papers in this and similar areas; some examples are \cite{brevik07,kovtun12}, and \cite{plumari12}.

Let us give a bit of the formalism. The fluid's energy-momentum tensor is
\begin{equation}
T_{\mu\nu}=\rho U_\mu U_\nu+(p-\zeta \theta)h_{\mu\nu}-2\eta \sigma_{\mu\nu}, \label{39}
\end{equation}
where $h_{\mu\nu}=g_{\mu\nu}+U_\mu U_\nu$ is the projection tensor, $\theta={U^\mu}_{;\mu}$ the scalar expansion, and $\sigma_{\mu\nu}=\theta_{\mu\nu}-\frac{1}{3}h_{\mu\nu}\theta$ with $\theta_{\mu\nu}=\frac{1}{2}(U_{\mu;\alpha}h_\nu^\alpha+U_{\nu;\alpha}h_\mu^\alpha)$ the shear stress tensor.

In principle, all the kinetic coefficients can, via the Kubo relations, be expressed as correlation functions of the corresponding currents. For the Fourier component $\eta(\omega)$ of $\eta$ the correlator is the stress tensor,
\begin{equation}
\eta(\omega)=\frac{1}{2\omega}\int dtdx e^{i\omega t}\langle [T_{xy}(t,x),T_{xy}(0,0)]\rangle; \label{40}
\end{equation}
(cf., for instance, \cite{hosoya84,brevik07}).

Our question is now: is the shear viscosity important in the early universe? It is here worth noticing that increasing attention is actually being paid to this kind of cosmology. Part of the reason for this has its roots in observations.  Recent WMAP measurements  indicate that the quadrupole and the octupole are aligned and concentrated in a plane about $30^o$ to the galactic plane, suggesting that there is an asymmetric expansion with one direction expanding differently from the other two directions (see, for instance,  \cite{tripathy14} with further references therein).

 It seems therefore worthwhile to investigate this case in some detail. We will take the anisotropic Kasner universe as a typical example,  corresponding to the metric
\begin{equation}
ds^2=-dt^2+t^{2p_1}dx^2+t^{2p_2}dy^2+t^{2p_3}dz^2, \label{41}
\end{equation}
the numbers $p_1, p_2, p_3$ being constants. One can introduce two numbers $P$ and $Q$ here, defined as $P=\sum_1^3 p_i, Q=\sum_1^3p_i^2$. In the usual Kasner theory for a vacuum, one has $P=Q=1$. This simple property no longer holds when the cosmic fluid possesses an energy density $\rho$, pressure $p$, and viscosity coefficients $\eta$ and $\zeta$, all of which are depending on time.

 The following theoretical description of  this situation is essentially extracted from  \cite{brevik04a}. From Einstein's equations, taking $\Lambda_4=0$, it is easy to obtain the following time-dependent solution
\[
\rho(t)=\rho_*\left(\frac{t_*}{t}\right)^2, \quad p(t)=p_*\left(\frac{t_*}{t}\right)^2, \]
\begin{equation}
\zeta(t) = \zeta_*\frac{t_*}{t}, \quad \eta(t)=\eta_*\frac{t_*}{t}, \label{42}
\end{equation}
where $\{ \rho_*, p_*, \zeta_*, \eta_*\}$ refer to some some chosen initial instant $t=t_*$. The Einstein equations then take the following convenient form involving time-independent quantities only,
\begin{equation}
P-Q+\frac{3}{2}\kappa_4^2\zeta_*t_* P=\frac{1}{2}\kappa_4^2t_*^2(\rho_*+3p_*), \label{43}
\end{equation}
\begin{equation}
p_i(1-P-2\kappa_4^2\eta_*t_*)+\frac{1}{2}\kappa_4^2t_*(\zeta_*+\frac{4}{3}\eta_*)P=-\frac{1}{2}\kappa_4^2t_*^2(\rho_*-p_*) \label{44}
\end{equation}
(recall that $\kappa_4^2=8\pi G_4$ in the present notation). From these equations $P$ and $Q$ can be found, once the $p_i$ are known.

We can now easily calculate the rate of entropy production. The average expansion anisotropy parameter is defined as
\begin{equation}
A=\frac{1}{3}\sum_{i=1}^3\left( 1-\frac{H_i}{H}\right)^2, \label{45}
\end{equation}
where $H_i=\dot{a}_i/a_i$ with $a_i=t^{p_i}$, $H=\frac{1}{3}\sum_1^3H_i$ being the average Hubble factor. This means that $A$ can be expressed as $A=3Q/P^2-1$. The entropy four-current is $S^\mu=n\sigma U^\mu$, with $n$ the baryon number density and $\sigma=s/n$ the nondimensional entropy per baryon. In the comoving frame of reference the rate of entropy production is thus
\begin{equation}
\dot{\sigma}=\frac{3P^2}{nTt^2}\left( \zeta +\frac{2}{3}\eta A \right). \label{46}
\end{equation}
This simple relationship enables us to determine the amount of anisotropy in the plasma era for which the shear and bulk viscosity are of the same importance. From Eq.~(\ref{46}) we have  for this case
\begin{equation}
A \approx \frac{3}{2}\frac{\zeta}{\eta}. \label{47}
\end{equation}
From the tabular data given in \cite{hogeveen86} we have at $T=10^{12}~$K (cgs units): $\eta =9.81\times 10^{23}, \, \zeta =2.80\times 10^{13}$, leading to $A\approx 4\times 10^{-11}$, an immensely small number. At the lower end of the era, for $T=5\times 10^9~$K, we have $\eta =2.02\times 10^{26}, \, \zeta =1.55\times 10^{24}$, implying $A\approx 1\times 10^{-2}$. There is thus a very large difference between the two cases as regards influence upon entropy production; the anisotropy parameter having to be much higher at the lowest temperatures in order to make the influence from $\eta$ appreciable.

\section{Summary}

Our analysis has been restricted to the plasma era only, i.e., the period for which the temperature drops from about $10^{12}~$K to $5\times 10^9~$K. In the earlier parts of this era we have found that the bulk viscosity $\zeta$ plays a negligible role. That includes the first order quark-gluon phase transition occurring at about $10^{12}~$K. In the later part of the era, the influence from viscosity may be discernible, a few per cent typically. Our analysis has been based upon the value for $\zeta(T)$ calculated from kinetic theory, to the first order in the inverse temperature expansion; cf. Eq.~(\ref{20}).  Equations (\ref{33}) and (\ref{34}) determine the relationship between  temperature $T=T(t)$ and time $t$, at times after the phase transition.

Although the magnitudes of the calculated viscosity corrections in the situation considered here are numerically small, the mathematical procedure as such is general, and may be applicable for other eras in the universe's history for which the effect can be greater.

Once anisotropies in the geometry of the universe are incorporated, the shear viscosity $\eta$ also comes into play. Section 3 discusses the  Kasner universe as an example.  Equation (\ref{47})  shows the magnitude of the anisotropy parameter $A$ for which  $\zeta$ and $\eta$ are of similar importance for the local entropy production. It turns out that the value of $A$ is relatively big, $A \approx 10^{-2}$, in the later part of the plasma era. Here the values of $\eta$, as well as $\zeta$, as functions of time, are adopted from kinetic theory.

\bigskip

{\bf Acknowledgements}

\bigskip

\noindent I thank Kazuo Ghoroku for valuable discussions and correspondence on this topic. Also, I thank Lars Husdal for making me aware of the paper by \cite{hogeveen86}.

\end{document}